# Evaluation of estimation approaches on the quality and robustness of collision warning system

Masoud Baghbahari
ECE Department and NanoScience Technology Center
University of Central Florida
Orlando, FL 32816, USA
bahari@knights.ucf.edu

Neda Hajiakhoond
Department of computer Science
University of Central Florida
Orlando, FL 32816, USA
Hajiakhoond@knights.ucf.edu

*Abstract* - **Vehicle safety is one of the most challenging aspect of future-generation autonomous and semi-autonomous vehicles. Collision warning systems (CCWs), as a proposed solution framework, can be relied as the main structure to address the issues in this area. In this framework, information plays a very important role. Each vehicle has access to its own information immediately. However, another vehicle information is available through a wireless communication. Data loss is very common issue for such communication approach. As a consequence, CCW would suffer from providing late or false detection awareness. Robust estimation of lost data is of this paper interest which its goal is to reconstruct or estimate lost network data from previous available or estimated data as close to actual values as possible under different rate of lost. In this paper, we will investigate and evaluate three different algorithms including constant velocity, constant acceleration and Kalman estimator for this purpose. We make a comparison between their performance which reveals the ability of them in term of accuracy and robustness for estimation and prediction based on previous samples which at the end affects the quality of CCW in awareness generation.**

## I. Introduction and Related Works

Safety is one of the primary and essential requirement for future intelligence systems such as transportation. Efforts in this area focus on reliable approaches which collision avoidance with other moving vehicles is their predefined objective. A lot of attempts have been conducted on vehicle safety challenge specifically in the form of cooperative vehicle safety. Among all of them, one of attractive concept is the problem of performance degradation due to network capacity saturation and transmitted data lost. Several algorithms aiming to face with this issue have been presented in literature [1]-[2]. Working on this issue has led to create two different attitudes in research publications. The first one has targeted network performance measure for enhancement which tries to reduce the packet error ratio (PER). Another perspective employs the estimation of vehicle position tracking error (PTE) as the performance criteria. Such views consider safety application as main objective rather than network performance. However, higher performance of cooperative collision warning systems is expected.
One of the most successful suggested algorithm in this regard is collision warning system. Vehicle-to-vehicle communication as the backbone of this system is responsible to convey vital data. This communication tool provides such flexibility for the system that every algorithm aiming in safety, cooperative control and monitoring can be implemented regardless of required data availability. In spite of this advantage, one inherent shortage appears as a new issue. Intrinsically for communication media, especially for wireless networks, data lost is still the most overwhelming matter. Basically, in this framework each vehicle as a separate node spreads its information thorough a wireless channel utilizing a very high data transmission known as Dedicated Short Range Communication (DSRC). Hundreds of meters is the coverage range for the broadcast data in the form of so called Basic Safety Message (BSM). Received data is employed in a real-time updated map to monitor and investigate other cars situation in terms of availability in collision potential zones. Depending on the type of vehicle receiving data, the action can be done automatically or intentionally by driver.

Regarding this fact and similar to [3] our working philosophy in this paper is to consider Forward Collision Warning (FCW) system as a special kind of cooperative collision warning systems. Instead of receiving data via sensor, laser or radar, BSM is accountable for information broadcasting. This crucial information are those that might be implemented in any data dependent procedure and also for next driving action. However, the responsive quality of any data-driven approach also depends on the availability and accuracy of required data. In networking specifically in such wireless networks, data lost is a very common issue [4]-[5]. For future real time applications such as Internet of Things [9]-[10], real time estimations approaches can strengthen the machine learning strategies in modeling and data mining based on online available data [11]-[12].

This paper is an effort to analyze and deal with this innate deficiency of losing data over wireless network on the FCW algorithm performance. After briefly explanation of collision warning system framework, several approaches will be presented and their performance will be investigated. To validate each presented algorithm, a large data set accumulated from hazard zone of 100 cars would be considered. The performance validation will be done on a data-set of driving situation in which last-second brake or steering were the action to escape from the hazardous conditions.

Section II is devoted to describe the collision warning system. Evaluation metrics and the estimation approaches are provided in section III. Simulation, evaluation and comparison between them is presented in section IV and finally section V describes the conclusion of this work.

## II. SYSTEM DESCRIPTION

CCW system consists of several subsystems. The interaction between these subsystems generates the desired output. All these subsystems would be describe in this section and the model of each components would be presented.

### A. System constituents

Basically in CCW systems, accessibility of information is via inter-vehicle broadcasting DSRC based wireless network. Transmitted data in this network are such informative that easily provides a comprehensive information about the relative situation of other vehicle.
Clearly, such sufficient information is a very excellent source for tracking the behavior of other near vehicles, aiming to do suitable and fast reaction. Received BSMs from the network decode to identify sending nodes. An instantaneous updating map tries to capture the situation of each nodes in awareness system. The assigned duty of collision predication algorithm is to analyze this map to forecast next possible behavior [3].

### B. Collision warning algorithm

FCW algorithms are responsible to detect feasible collisions. The outcome of this detection is alteration of driving system or driver reaction. To propose an efficient algorithm, some considerations are needed. The most problematic part is that it is critical to generate an alarm with correct timing [3]. Among all suggested algorithms for this goal our selection is so called CAMPLinear algorithm [6]-[7]. This algorithm is based on a huge volume of data sets of driver's reaction. The main idea behind of this approach is monitoring the distance between two vehicles constantly. These two vehicle are known as Leading Vehicle (LV) and Following Vehicle (FV). As long as they are moving in a predefined hazardous distance threshold, a collision warning signal will be generated. The relative distance between any two moving vehicles is related to their relative speeds. Based on algorithm explanation and mathematical notation from [3], the warning range between two cars $r_w$ which is the summation of Brake Onset Range (BOR) and operator reaction $r_d$ govern by following equation:

$$r_w = BOR + r_d \quad (1)$$

$$r_d = 0.5(a_{FV} - a_{LV})t_d^2 + (v_{FV} - v_{LV})t_d \quad (2)$$

In which $t_d$ is the required time for operator and brake system reaction.
It is also worth noting that the second equation is so called kinematic equation relating distance to acceleration and speed according to relationship between distance with acceleration and velocity in time domain. Three different scenarios are possible to compute BOR as follow [3]:
Fist one is stationary state for LV at the beginning and at end in which we have:

$$BOR_1 = -\frac{V_{FVP}^2}{2d_{rqd}} \quad (3)$$

In second scenario, moving is the state of LV at the beginning and at the end:

$$BOR_2 = -\frac{(v_{FVP} - v_{LVP})^2}{-2(d_{rqd} - d_{LV})} \quad (4)$$

The last one would be moving at the beginning and stopping at the end:

$$BOR_3 = \frac{v_{FVP}^2}{-2d_{rdq}} - \frac{v_{LVP}^2}{-2d_{LV}} \quad (5)$$

In these equations $v_{FVP}$ and $v_{LVP}$ are indication of the predicted speed of FV and LV. We can use the dependency of each vehicle acceleration $a_v$ to its speed $v_v$ according to first order kinematic equation to find the prediction during $t_d$ elapsed time:

$$v_{FVP} = v_{FV} + a_{FV}t_d \quad (6)$$

$$v_{LVP} = v_{LV} + a_{LV}t_d \quad (7)$$

The primary assumption for these equations is constant acceleration during $t_d$. The terms $d_{LV}$ is LV deceleration and $d_{rqd}$ required FV deceleration for crash avoidance can be modeled [6]-[7]:

$$d_{rqd} = -5.3 + 0.68a_{Lv} + 2.57(v_{LV} > 0) \\ - 0.086(v_{FV} - v_{LVP}) \quad (8)$$

Obviously exact information of FV is always available for itself, however, LV state information is accessible through network. In such case, we need to use the speed estimation $v_{LV}$ and $a_{LV}$ instead of actual value. To put in perspective, we can simplify the system into a block diagram such as Fig. 1.

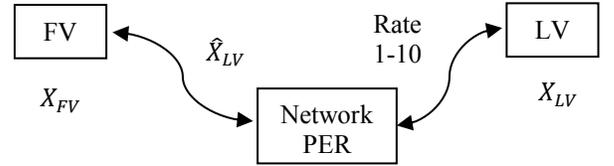

Fig. 1 CCW System Block Diagram

The estimation process can be done by different techniques. In next section under different assumptions, we will attempt to extract estimated state information $X_{LV}$ from received data which is the required input for CAMPLinear algorithm.

## III. PERFORMANCE EVALUATION AND ESTIMATION APPROACHES

### A. Performance Evaluation

The FCW algorithm is a real-time approach. At each instant of time, a running process constantly investigates and monitors current situation between two vehicles $d(t)$, continuously compares it with warning threshold distance $r_{w(t)}$ to generate notification about upcoming unsafe situations. To successfully accomplish this task, continuously access to exact state information of two vehicles are vital. However, the media of sending and receiving data in this case is a wireless network. Dropping part of receiving data during transmission process or delay in receiving data through the channel is very natural.

To handle this intrinsic drawback, extraction or estimation of next step possible states from previous received or estimated samples is a suggested solution. Before explaining the estimation frameworks, we need some criteria to validate frameworks performance in collision warning detection and to make a fair comparison between them.

Performance metrics are responsible to express performance measure quantitatively. Performance measurement can be achieved using performance metrics. A wide variety of methods provided in [8] are a noticeable performance measures suitable for collision avoidance approaches. Among all of them, our choice for evaluation in this paper are accuracy and true positive which obtain by following simple formulas [similar to [8]]:

$$TP = \frac{Ch}{Is + Ch} \quad (9)$$

$$Accu = \frac{Ch + Cs}{Is + Ih + Ch + Cs} \quad (10)$$

In which $Ch$ is the number of correct hazard predictions and $Cs$ is the number of safe indications, also $Is$ indicates the number of incorrect safe indications predication and $Ih$ is the number of incorrect hazard prediction. According to accuracy equation it just provides the overall accuracy of system as the portion of total of generated and no generated warnings to the whole possible generations [8].

A. *Estimation Approaches*

Kinematics equation is an equation expressing the time domain relationship between two variables. The first order kinematics stating the relation between velocity ( $v$ ) and position ($x$) is:

$$v = \frac{dx(t)}{dt} \quad (11)$$

And for acceleration ($a$) and velocity:

$$a = \frac{dv(t)}{dt} \quad (12)$$

The second order kinematics equation eliminates the velocity form these equations to express the direct relationship between acceleration and position directly:

$$a = \frac{dx^2(t)}{dt^2} \quad (13)$$

If the velocity be constant during an elapsed time $t_d$, from the first order kinematics one can obtain the position in next sample time using previous sample:

$$x(t + t_d) = x(t) + vt_d \quad (14)$$

For FCW the elapsed time from previous sample depends on the frequency which the algorithm works accordingly and can be vary form $100ms$ ($10Hz$) to $1s$ ($1Hz$). Similarly for the velocity if the acceleration during elapsed time be constant, the next step velocity can be achieved from previous one by:

$$v(t + t_d) = v(t) + at_d \quad (15)$$

Using these equations the whole process for FCW can be explained. As long as the algorithm receives required state (acceleration, velocity and position) data from LV via wireless network, it can run the detection process with such data. On the contrary, if data be dropped or corrupted during transmission, an estimation of states from equations (14) and (15) will be implemented. Clearly the assumptions used by these equations have to be satisfied to have a close prediction from previous received sample data. It is also possible to consider the direct relationship between position and acceleration. In this case one can propose Kalman filter as a choice to have estimation from previous samples. One of the most important feature of Kalman filter which can make it a good choice for this case is in its ability to track the signal despite of noise and uncertainty. As a matter of fact, the design of its gains is under the assumption of process noisy information. It also will be shown by simulation results, this unique property of filter provides better results by performing more robust against the lost information as a source of uncertainty specifically in higher value of dropping packets rate. The basic differential equation which to be used here is the second order kinematics equation which express the relationship between acceleration and position. The transfer function in frequency ($s$) domain is:

$$x(s) = \frac{a(s)}{s^2} \quad (16)$$

To implement Kalman filter successfully, the dynamic equation should convert to state space model in time ($t$) domain:

$$\frac{dX(t)}{dt} = AX(t) + Bu(t) \quad (17)$$
$$y(t) = CX(t) + Du(t)$$

In which $u(t)$ and $y(t)$ are the dynamic input and output signals respectively and in this case input signal would be the received acceleration and output signal is the position. $X(t)$ is the system state vector. This vector captures the states of system which are the velocity and position.

$$X(t) = \begin{bmatrix} v(t) \\ x(t) \end{bmatrix} \quad (18)$$

For the relationship between acceleration and position (16) the corresponding state space matrices in (17) can be expressed as:

$$A = \begin{bmatrix} 0 & 0 \\ 1 & 0 \end{bmatrix}, B = \begin{bmatrix} 1 \\ 0 \end{bmatrix}, C = \begin{bmatrix} 0 & 1 \end{bmatrix}, D = 0 \quad (19)$$

Using these matrices it is straightforward to implement Kalman filter by Matlab Control Toolbox. In next section, these three suggested estimation approaches are implemented and their evaluation in terms of true positive and accuracy alongside the comparison between them would be presented.

## IV. SIMULATION

To evaluate and compare the performance of proposed estimation frameworks, we use a rich data-set of 100 scenarios trajectories similar to [3]. The reason behind the richness of this data-set is that some scenarios belong to near crash or actual crash scenarios. As a result, it can be an excellent data-set reference to evaluate and validate of proposed crash warning algorithms. Here by selection of CAMP Linear as crash warning algorithm, we implement three previous section proposed estimation approaches for LV state tracking. The ability of such methods is analyzed in terms of accuracy and true positive performance metric.

The sampling and communication frequency which we use for simulation is $10Hz$ and it is equal to $100ms$ sampling time.
First we fix PER on 0.3 which means loss information happen for 3 out of 10 packets. If we implement constant velocity estimation method, the result indicating the difference between actual velocity and estimated velocity is highlighted in Fig. 2.

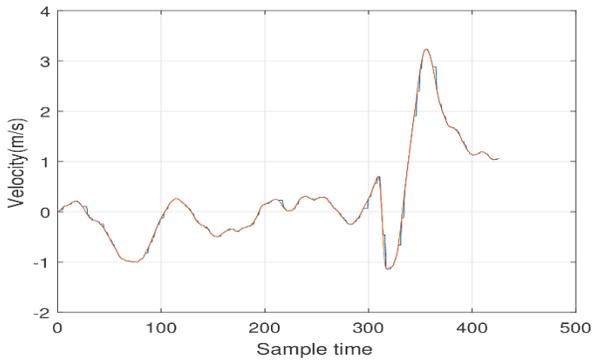
Fig. 2 Actual Velocity vs Estimated Velocity

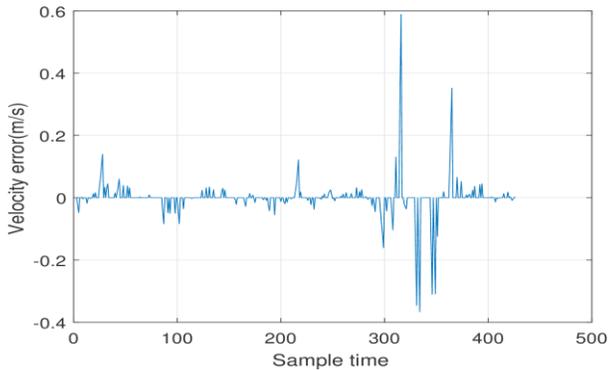
Fig. 3 Error between Actual Velocity and Estimated Velocity

To have a better insight about the difference between actual and estimated velocity Fig. 3 shows the error between these values. Using equation (14) we can compute estimated position which is shown it in Fig. 4.

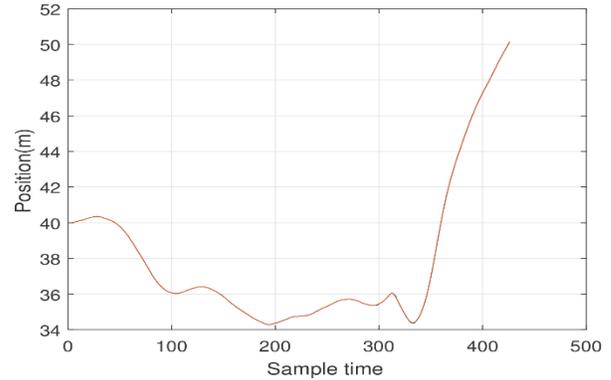
Fig. 4 Actual Position vs Estimated Position

And the error between actual position and estimated position can be presented easily in Fig. 5. From this figure it is obvious that estimated position and actual position are so close to each other. Similarly from (15), the acceleration can be shown in Fig. 6.

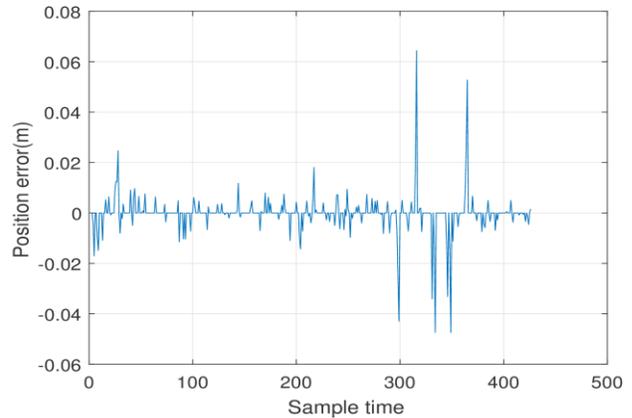
Fig. 5 Error between Actual Position and Estimated Position

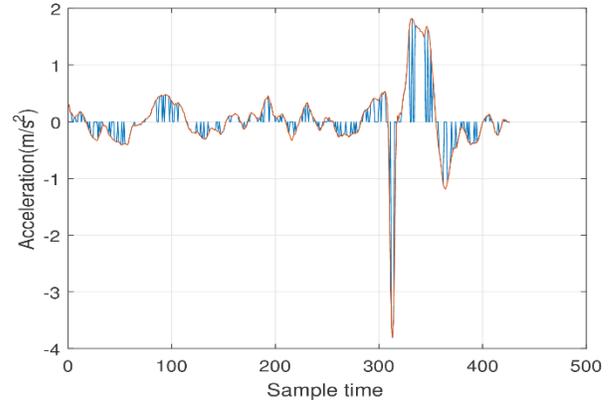
Fig. 6 Actual Acceleration vs Estimated Acceleration

Several jumps from acceleration to zero value are due to constant velocity assumption during loss data time. Such assumption simply means zero acceleration. The acceleration error can be seen in Fig. 7.

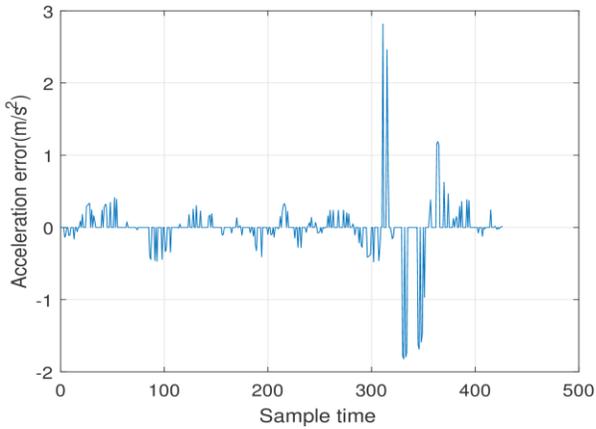

Fig. 7 Error between Actual Acceleration and Estimated Acceleration

Now if we consider constant acceleration model form (15), the error between actual and estimated LV states including velocity, position and acceleration represent in Fig. 8, Fig. 9, and Fig. 10 respectively, which shows better estimation for LV state information.

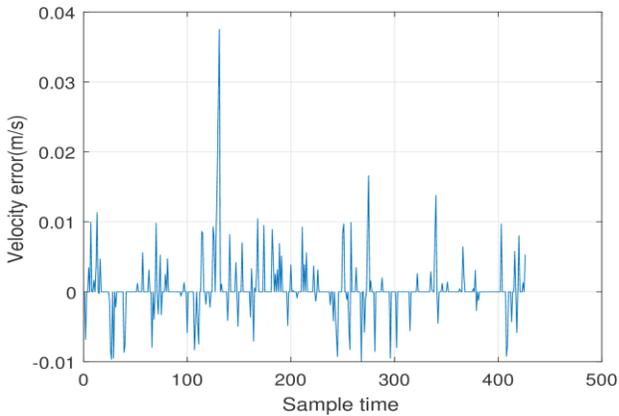

Fig. 8 Error between Actual Velocity and Estimated Velocity

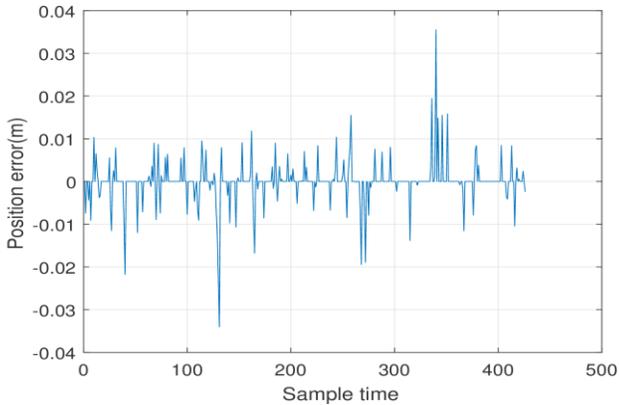

Fig. 9 Error between Actual Position and Estimated Position

The results for Kalman filter implementation can be observed from Fig. 11, Fig. 12 and Fig. 13.

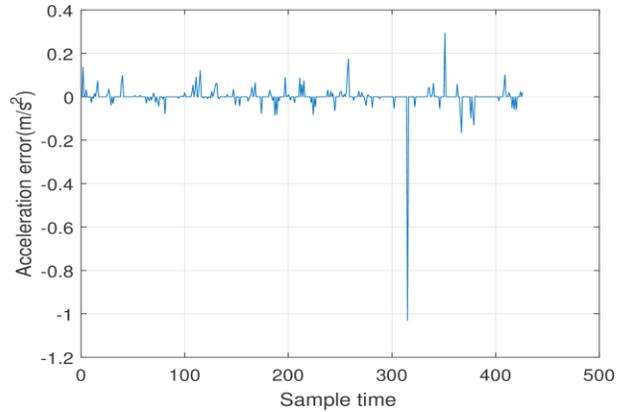

Fig. 10 Error between Actual Acceleration and Estimated Acceleration

In next step we consider the effect of network data loss by changing the PER value from 0.1 to 0.9. Since the availability of exact data for estimation algorithms reduce by higher value of PER, degradation of estimation from actual value because of lack of new data is unavoidable. Aiming to evaluate the FCW algorithm under different PER and different estimation frameworks, we would present the average value of true positive and accuracy for 100 car scenarios.

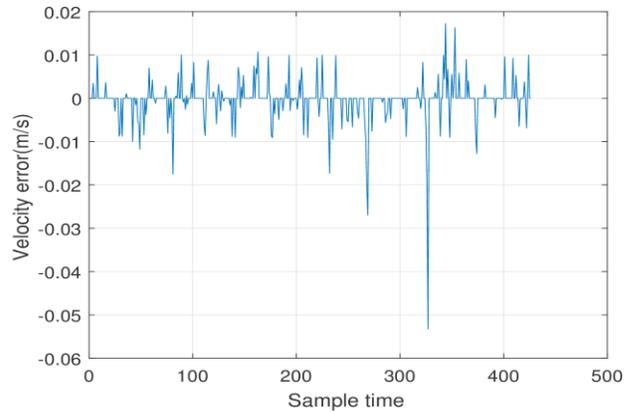

Fig. 11 Error between Actual Velocity and Estimated Velocity

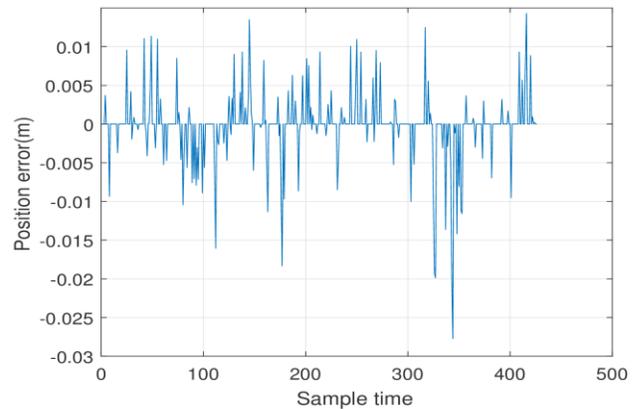

Fig. 12 Error between Actual Position and Estimated Position

The accuracy and true positive for constant velocity model can be seen in Fig. 14 and Fig. 15. Naturally, when the value of PER

is low, the estimation will be closer to real values and as a consequence the output of algorithm warning would be more precise.

Constant acceleration method provides better results rather than constant velocity as Fig. 16 and Fig.17 also confirm this explanation. So the constant acceleration assumption for vehicles seems closer to reality than constant velocity.

The results for Kalman filter is more noticeable. From Fig. 18 and Fig. 19, for lower value of PER, the accuracy and true positive are better than constant velocity and worse than constant acceleration model, however, for higher value of PER the result is a bit better than both constant velocity and acceleration. So it means that Kalman filter performs a bit more robust when the availability of data reduces as a result of packet dropping and higher value of PER.

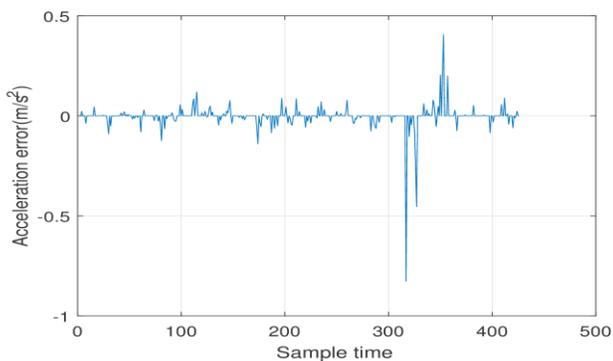
Fig. 13 Error between Actual Acceleration and Estimated Acceleration

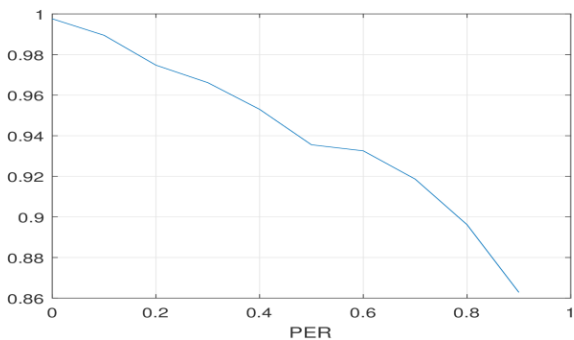
Fig. 14 accuracy vs PER for Constant Velocity Estimation

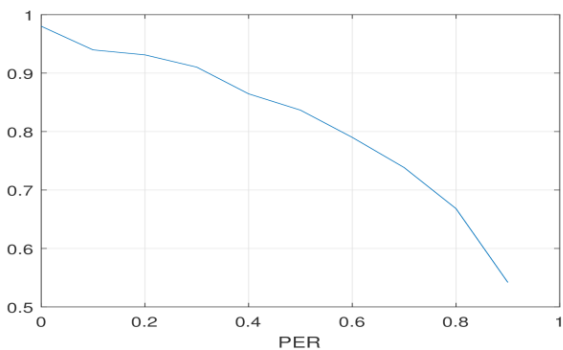
Fig. 15 True Positive vs PER for Constant Velocity Estimation

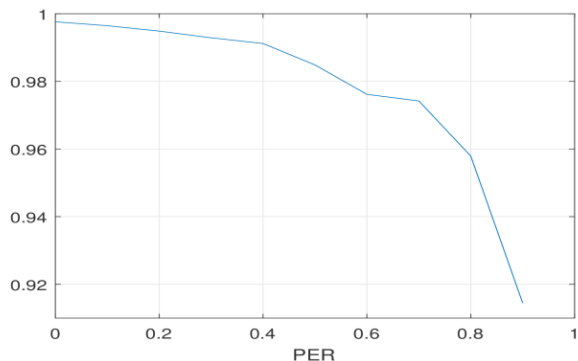
Fig. 16 accuracy vs PER for Constant Acceleration Estimation

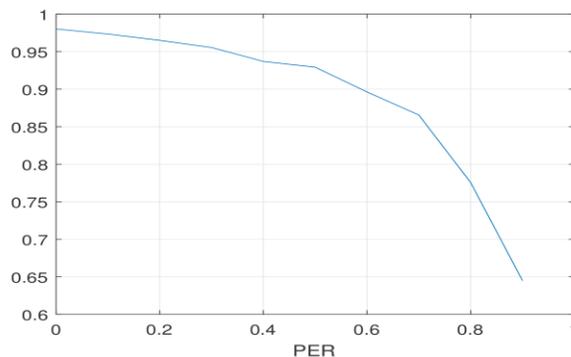
Fig. 17 True Positive vs PER for Constant Acceleration Estimation

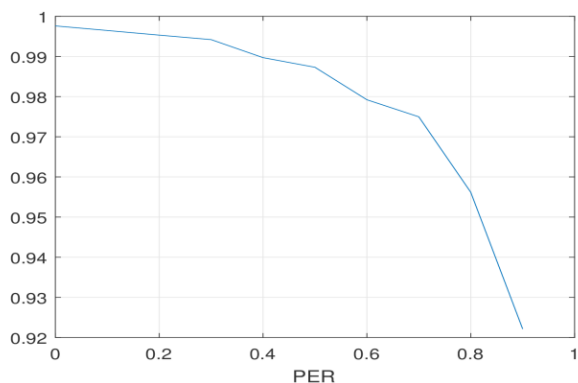
Fig. 18 Accuracy vs PER for Kalman Filter Estimation

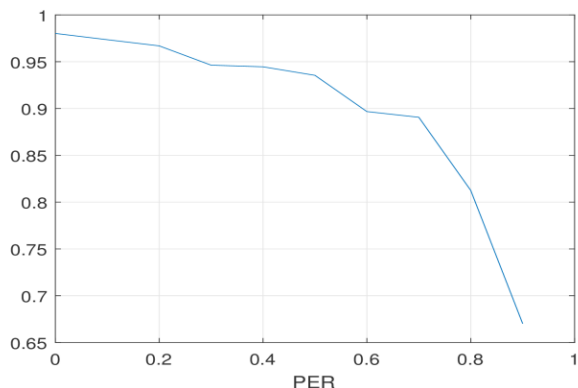
Fig. 19 True Positive vs PER for Kalman Filter Estimation

## V. Conclusion

This paper is an effort to describe and analyze the collision warning system as a noticeable approach for safety of next-generation of autonomous vehicles under different estimation implementation. Having precise information from remote vehicle helps the algorithm to provide more promising outcomes. Due to lost data during transmission process, we investigate the robustness and performance of three different estimation frameworks. Simulation results from a rich data set of car trajectories shows the effectiveness of each of them with different PER in terms of accuracy and true positive evaluation metrics.